\documentclass[reprint, amsmath,amssymb, aps, prb, superscriptaddress]{revtex4-2}
\usepackage{graphicx}
\usepackage{dcolumn}
\usepackage{bm}
\usepackage[colorlinks=true, citecolor=blue]{hyperref}
\usepackage{dcolumn}
\usepackage{cleveref}
\usepackage{color}
\usepackage{multirow}
\usepackage{float}
\usepackage{url}
\usepackage{physics}
\usepackage{subfigure}
\usepackage[utf8]{inputenc}
\usepackage{siunitx}
\usepackage{ragged2e}
\usepackage{titlesec}
\usepackage[T1]{fontenc}
\usepackage[dvipsnames]{xcolor}
\usepackage[version=3]{mhchem}
\graphicspath{ {./Figures/} }

\begin{document}

\title{Possible Proximity to Ferromagnetism in the \ce{V2Ga5} Superconductor}

\author{Szymon Kr\'{o}lak}
\email{szymon.krolak@pg.edu.pl}
\affiliation{Faculty of Applied Physics and Mathematics and Advanced Materials Center, Gdansk University of Technology, Narutowicza 11/12, 80-233 Gdansk, Poland}

\author{Xudong Huai}
\affiliation{Department of Chemistry, Clemson University, Clemson, South Carolina 29634, United States}

\author{Wiktoria Jarosz}
\affiliation{Faculty of Applied Physics and Mathematics and Advanced Materials Center, Gdansk University of Technology, Narutowicza 11/12, 80-233 Gdansk, Poland}

\author{Filip Ko\v{s}uth}
\affiliation{Centre of Low Temperature Physics, Institute of Experimental Physics, Slovak Academy of Sciences, SK-04001, Košice, Slovakia}
\affiliation{Centre of Low Temperature Physics, Faculty of Science, P.J. Šafárik University, SK-04001, Košice, Slovakia}

\author{Pavol Szab\'{o}}
\affiliation{Centre of Low Temperature Physics, Institute of Experimental Physics, Slovak Academy of Sciences, SK-04001, Košice, Slovakia}

\author{Micha\l{} J. Winiarski}
\affiliation{Faculty of Applied Physics and Mathematics and Advanced Materials Center, Gdansk University of Technology, Narutowicza 11/12, 80-233 Gdansk, Poland}

\author{Sudip Malick}
\affiliation{Faculty of Applied Physics and Mathematics and Advanced Materials Center, Gdansk University of Technology, Narutowicza 11/12, 80-233 Gdansk, Poland}

\author{Thao T. Tran}
\affiliation{Department of Chemistry, Clemson University, Clemson, South Carolina 29634, United States}

\author{Tomasz Klimczuk}
\email{tomasz.klimczuk@pg.edu.pl}
\affiliation{Faculty of Applied Physics and Mathematics and Advanced Materials Center, Gdansk University of Technology, Narutowicza 11/12, 80-233 Gdansk, Poland}

\begin{abstract}
Superconductivity and ferromagnetism are generally competing ground states in $d$-electron systems, making their interplay of fundamental interest. We report a comprehensive study of high-quality single- and polycrystalline \ce{V2Ga5}, a bulk type-II superconductor ($T_{\mathrm{c}} = 3.54$ K) with a quasi-one-dimensional crystal structure, supplemented with density functional theory (DFT) calculations, suggesting possible proximity to ferromagnetic order. Below $T \approx 10$ K, magnetic susceptibility shows ZFC/FC splitting, along with saturation and hysteresis in $M(H)$. Moreover, electrical transport measurements reveal a magnetic-field-dependent resistivity upturn, while specific heat is enhanced in magnetic fields. DFT calculations show that the Fermi level in \ce{V2Ga5} is located at a peak in the density of states, with a small magnetic moment per unit cell comparable to the experimental value. Together, these results indicate the possibility that ferromagnetic correlations develop below $T \approx 10$ K, well above $T_{\mathrm{c}}$, with long-range ferromagnetic order suppressed by the superconducting transition.
\end{abstract}

\maketitle

\section{Introduction}

The interplay between ferromagnetism (FM) and superconductivity (SC) has long attracted interest, as the two phenomena are generally considered antagonistic due to the pair-breaking effects of internal exchange fields associated with ferromagnetic order. Well-known exceptions occur in certain \textit{f}-electron systems: \ce{UGe2}, \ce{UCoGe}, and \ce{URhGe} \cite{UGe2, UCoGe, URhGe}, where SC coexists with FM. These compounds, characterized by localized \textit{f}-electrons, exhibit spin-triplet pairing, which permits the coexistence of the two orders. Comprehensive reviews of this class of materials can be found in Refs. \cite{Huxley2015, Aoki2011}.

In contrast, electrons in \textit{d}-electron systems are considerably more itinerant, and superconductivity generally competes rather than coexists with ferromagnetism. Both phenomena originate from instabilities of the electronic system, with ferromagnetism prevailing when the Stoner criterion is satisfied \cite{Stoner}. Although coexistence of SC and FM has been reported in \textit{d}-electron compounds such as \ce{ZrZn2} \cite{ZrZn2-nature} and \ce{Y9Co7} (previously identified as \ce{Y4Co3}) \cite{Y9Co7-first}, these cases remain controversial. In the itinerant ferromagnet \ce{ZrZn2}, superconductivity was later shown to originate solely from a surface layer \cite{ZrZn2-spark}, while in \ce{Y9Co7}, ferromagnetism has not been conclusively established, partly due to strong sample dependence of physical properties and the absence of studies on single crystals \cite{Huxley2015}. Thus, despite several intriguing reports, conclusive evidence for the coexistence of superconductivity and ferromagnetism in \textit{d}-electron systems is still lacking.

For understanding the interplay between ferromagnetism and superconductivity, the symmetry of the superconducting order parameter is a key factor. In U-based compounds, the coexistence of FM and SC ground states is well explained within the framework of spin-triplet pairing, which is inherently robust against internal magnetic fields \cite{Huxley2015, Aoki2011}. In contrast, conventional spin-singlet \textit{s}-wave superconductivity is usually expected to be strongly suppressed by the exchange field associated with ferromagnetic order \cite{Sarma1963}. In this context, the finding of superconductivity in \ce{MgCNi3} sparked significant interest~\cite{MgCNi3-Cava}. Due to its large density of states at the Fermi level, DOS($E_{\mathrm{F}}$), \ce{MgCNi3} was considered a promising candidate for the emergence of ferromagnetism via hole (e.g. Co, Ru) doping, by shifting the Fermi level to satisfy the Stoner criterion~\cite{MgCNi3-magnet}. However, despite these theoretical expectations, experimental efforts to induce ferromagnetism through hole doping have been largely unsuccessful~\cite{Klimczuk2004-ru, Klimczuk2004-fu-ru, Klimczuk2005-b, Das2003, Kumary2002}. As a result, the interplay between superconductivity and ferromagnetism in this system remains unexplored. Subsequent measurements on high-quality single crystals established that \ce{MgCNi3} is a conventional, \textit{s}-wave superconductor \cite{Gordon2013, Pribulova2011, Diener2009, Lee2008}, ruling out spin-triplet pairing of the type observed in U-based systems.

The present case differs in an essential way. \ce{Mn2Ga5} is an established ferromagnet \cite{Mn2Ga5}, and Mn substitution is readily achieved in \ce{V2Ga5} \cite{Mn-V2Ga5}, indicating that, unlike \ce{MgCNi3}, ferromagnetism is \textit{experimentally} accessible. Mn-doping can be viewed, in a purely Stoner framework, as increasing DOS($E_{\mathrm{F}}$) until the Stoner criterion is met. From a chemical bonding perspective, however, the relevant effect is that doping drives the Fermi level deeper into V-V antibonding states. The occupation of antibonding orbitals was previously identified by Landrum and Dronskowski \cite{Landrum2000} as a chemical driving force for itinerant ferromagnetism, which was later confirmed experimentally \cite{Fokwa2007}. Thus, the fact that the Fermi level in \ce{V2Ga5} lies at the base of a large antibonding V-V DOS peak \cite{Rogacki-V2Ga5} suggests that \ce{V2Ga5} may be close to a ferromagnetic instability. Moreover, \ce{V2Ga5} is intrinsically free of magnetic elements, unlike \ce{Y9Co7} and \ce{MgCNi3}, which enables better control of magnetic impurity phases. Finally, the quasi-one-dimensional nature of its electronic states enhances quantum fluctuations \cite{QC-dimension}, rendering the resulting magnetic order especially fragile. This combination of tunable electronic structure, absence of magnetic elements, and reduced dimensionality makes \ce{V2Ga5} a promising compound to study the interplay of superconductivity and itinerant, $d$-electron ferromagnetism.

Although discovered decades ago \cite{REDDY1965}, \ce{V2Ga5} has only recently attracted renewed attention due to its potential topological superconductivity \cite{PRB-2024, PRR-2024}, supported by the observation of de Haas–van Alphen (dHvA) oscillations corresponding to bands with very light carriers ($m = 0.159 \ m_0$) \cite{PRB-2025}. Earlier dHvA measurements \cite{V2Ga5-Fermi}, together with angle-resolved photoemission spectroscopy (ARPES) \cite{PRR-2024}, confirmed the quasi-one-dimensional character of the Fermi surface. Muon spin rotation (\(\mu\)SR) and nuclear magnetic resonance (NMR) measurements \cite{scirep-2025} revealed pronounced electronic anisotropy associated with the quasi-1D vanadium chains and showed no evidence for static magnetic order either above or below $T_{\mathrm{c}}$. Consistent with these observations, specific heat measurements suggested conventional, two-gap \textit{s}-wave pairing \cite{PRB-2024, PRR-2024}, a conclusion further supported by \(\mu\)SR results \cite{PRR-2024, scirep-2025}.

In this work, we extend previous studies with a comprehensive investigation of \ce{V2Ga5} using magnetic susceptibility, electrical resistivity, heat capacity, and scanning tunneling microscopy (STM), complemented by DFT calculations. Our results show anomalous low-temperature behavior, possibly indicating the onset of ferromagnetism. DFT calculations support this interpretation, revealing that at $T = 0$ K \ce{V2Ga5} supports a magnetic moment, consistent with ferromagnetic ground state. The experimentally observed anomalies are reproducible in both single- and polycrystalline, high-quality samples from multiple synthesis batches, suggesting their intrinsic origin.

\section{Methods}

\subsection{Synthesis}

Single crystals of \ce{V2Ga5} were grown in \ce{Al2O3} crucibles using V powder (Alfa Aesar, 99.5\%) and Ga pieces (Onyxmet, 99.99\%), with a 6:94 V:Ga molar ratio. The evacuated quartz tube was heated to \SI{1000}{\degreeCelsius} at a rate of \SI{100}{\degreeCelsius\per\hour}, held at this temperature for \SI{48}{\hour}, and subsequently cooled to \SI{550}{\degreeCelsius} at a rate of \SI{2.5}{\degreeCelsius\per\hour}. At this temperature,
tubes were centrifuged to remove excess Ga flux. The resulting crystals were etched in lightly diluted hydrochloric acid (HCl) to eliminate residual Ga on the surface, following the procedure described in Ref. \cite{PRR-2024}. All crystals were synthesized under identical heating and cooling conditions, except for batch \#4, which was cooled at a rate of \SI{1}{\degreeCelsius\per\hour}.  

Polycrystalline \ce{V2Ga5} was synthesized from stoichiometric amounts of the same V and Ga precursors as used for the crystal growth, with the V powder pressed into pellets and arc-melted to form solid pieces prior to reaction with Ga. The synthesis was carried out in an Ar-filled chamber equipped with a water-cooled copper hearth, with Zr serving as a getter. Sample button was remelted several times to ensure homogeneity and subsequently annealed at $700^{\circ}\mathrm{C}$ for 5 days. The annealing step was found to be crucial for obtaining phase-pure samples.

\subsection{Physical property measurements}

\textbf{Crystal structure and chemical composition} 
Single-crystal diffraction experiments were performed on \ce{V2Ga5} crystals using a Bruker D8 QUEST diffractometer with Mo K$\alpha$ radiation ($\lambda = 0.71073$ \AA) and a PHOTON II 7 detector at $T = 300$ K. Data processing (SAINT) and scaling (SADABS) were carried out using the Apex4 software. The structure was solved by the intrinsic phasing method (SHELXT) and refined by full-matrix least-squares techniques on $F^{2}$ (SHELXL) using the SHELXL software suite. All atoms were refined anisotropically.

Powder X-ray diffraction was performed on ground \ce{V2Ga5} crystals with a Bruker D2 Phaser diffractometer using Cu K\(\alpha\) radiation (\( \lambda = \SI{1.5406}{\angstrom} \)) and a position-sensitive XE-T detector. The diffraction patterns were refined using the Rietveld method implemented in GSAS-II \cite{GSAS}. 

The chemical composition was examined by energy-dispersive X-ray (EDX) spectroscopy using a FEI Quanta FEG 250 electron microscope.

\textbf{Magnetization, resistivity, and heat capacity} 
All physical property measurements were performed using Quantum Design Physical Property Measurement Systems (PPMS, EverCool-II and Dynacool). Magnetic properties were measured with the vibrating sample magnetometer (VSM) option, with \ce{V2Ga5} crystals glued to the quartz holder with GE varnish. Electrical resistivity was determined using the conventional four-probe technique, with platinum wires attached to the crystal surface using a two-component silver epoxy. Specific heat was measured employing the two-tau relaxation method.

\textbf{Scanning Tunneling Microscopy}

The scanning tunneling microscopy (STM) experiments were performed using a homemade STM head in Košice, inserted in a commercial Janis SSV cryomagnetic system with a $^3$He refrigerator and controlled by Nanotec’s Dulcinea SPM electronics.

\subsection{Electronic structure calculations}

Density functional theory spin-polarized electronic structure calculations were performed using the ELK ver. 10.5.16 full potential linearized augmented plane wave (FP-LAPW) code \cite{elk} with the Perdew-Burke-Ernzerhof Generalized Gradient Approximation (PBE GGA) \cite{Perdew1996} of the exchange-correlation functional and with the spin-orbit coupling included. Brillouin zone integrations were completed on a 5x5x17 Monkhorst-pack k-point mesh. The DFT+U method was applied with the Around Mean Field double counting scheme using a range of U and J values. Cell parameters and atomic positions were based on the relaxed crystal structure taken from the Materials Project database (mp-20405) \cite{Jain2013,osti_1195532}

\section{Results}

\subsection{Crystal structure and chemical composition}

\begin{figure}[htbp]
\centering \includegraphics[width=0.99\columnwidth]{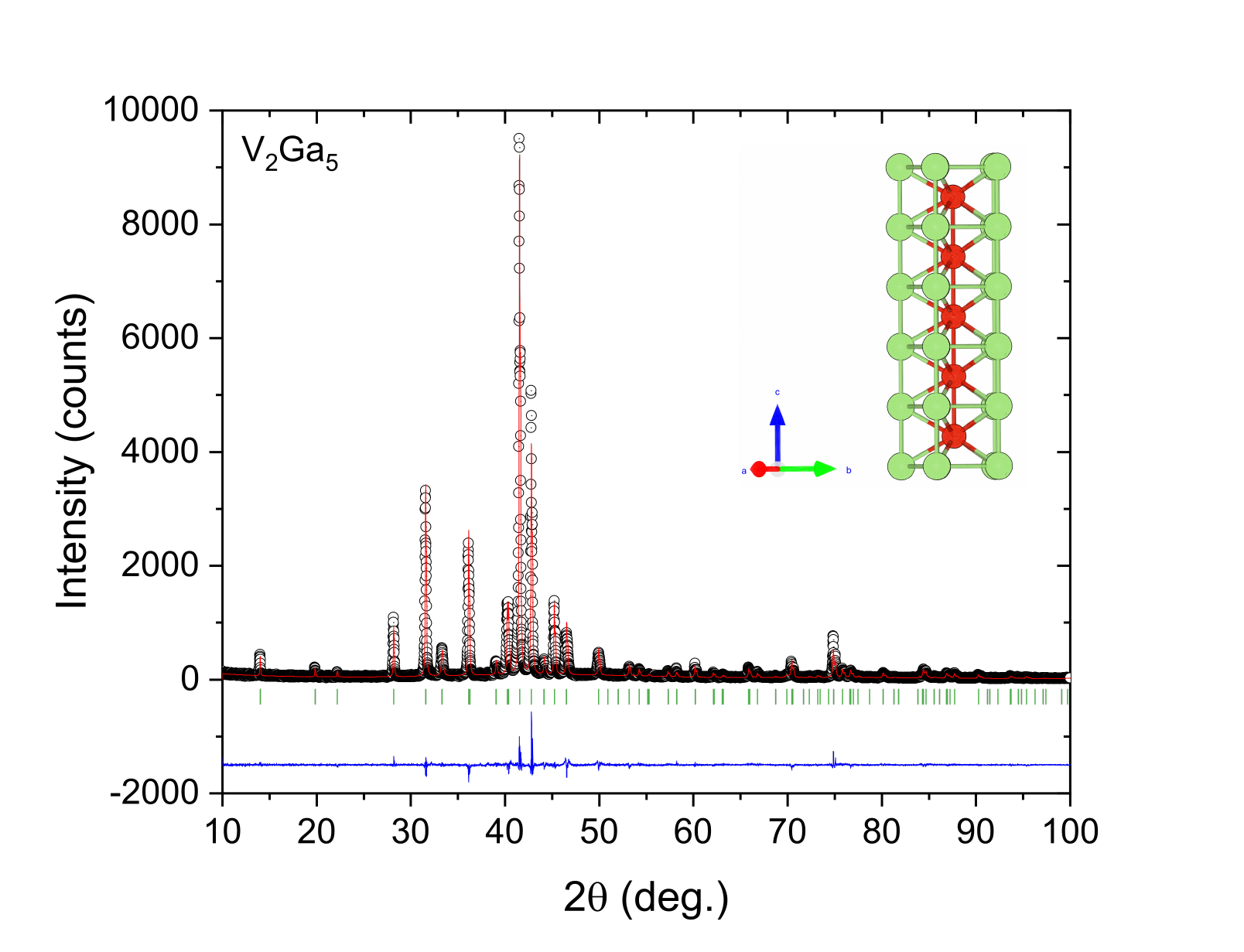}
\caption{Powder X-ray diffraction pattern of crushed \ce{V2G5} single crystals. Black circles represent measured data, the red line is the Rietveld refinement fit, green ticks indicate calculated Bragg peak positions, and the blue line is the difference curve. The inset illustrates the P4/mbm (s.g. no 127) crystal structure model used in the refinement, with Ga and V atoms represented by green and red balls, respectively.
}
\label{Fig:XRD}
\end{figure}

The quality of the \ce{V2Ga5} single crystals was first characterized with X-ray diffraction. Due to significant twinning, first suggested by the STM measurements (see Figure S1a in Supplementary Information), we performed single-crystal X-ray diffraction on a small piece of a crystal corresponding to a single twin, which allowed us to accurately determine the crystal structure (Table S1). Our results confirm the reported structure and are consistent with previous single-crystal diffraction measurements \cite{PRB-2024, PRR-2024, Rogacki-V2Ga5}. In fact, twinning appears to be characteristic of \ce{V2Ga5} single crystals, see also \cite{PRR-2024}. Using this structural model, we indexed the powder X-ray diffraction (pXRD) pattern collected on crushed \ce{V2Ga5} single crystals, which is presented in \Cref{Fig:XRD}, along with the results of the Rietveld analysis. All diffraction peaks are well indexed by the tetragonal \textit{P}4/\textit{mbm} space group (no. 127) with refined lattice parameters  a = 8.9643(1) \AA \ and c = 2.69028(6) \AA, in agreement with earlier reports \cite{REDDY1965, PRB-2024, PRR-2024, Rogacki-V2Ga5, V2Ga5-Fermi}.

Given the critical importance of precise stoichiometry and the absence of any impurity phases, standardized energy-dispersive X-ray spectroscopy (EDX) analysis on \ce{V2Ga5} single crystals was performed. A detailed description of the methodology and results is provided in the Supplementary Information. In brief, the data, summarized in Figure S1, show that the experimental stoichiometry is very close to the ideal 2:5 composition. Considering the experimental uncertainty of the EDX measurements and the fact that the site occupancies of V and Ga, as determined from Rietveld refinement (Table S2), are close to unity, we conclude that the \ce{V2Ga5} samples studied here are stoichiometric. Thus, we provide a comprehensive proof for both phase and chemical purity of \ce{V2Ga5} single crystals.

\subsection{Magnetic properties}

\begin{figure}[htbp]
\centering \includegraphics[width=0.99\columnwidth]{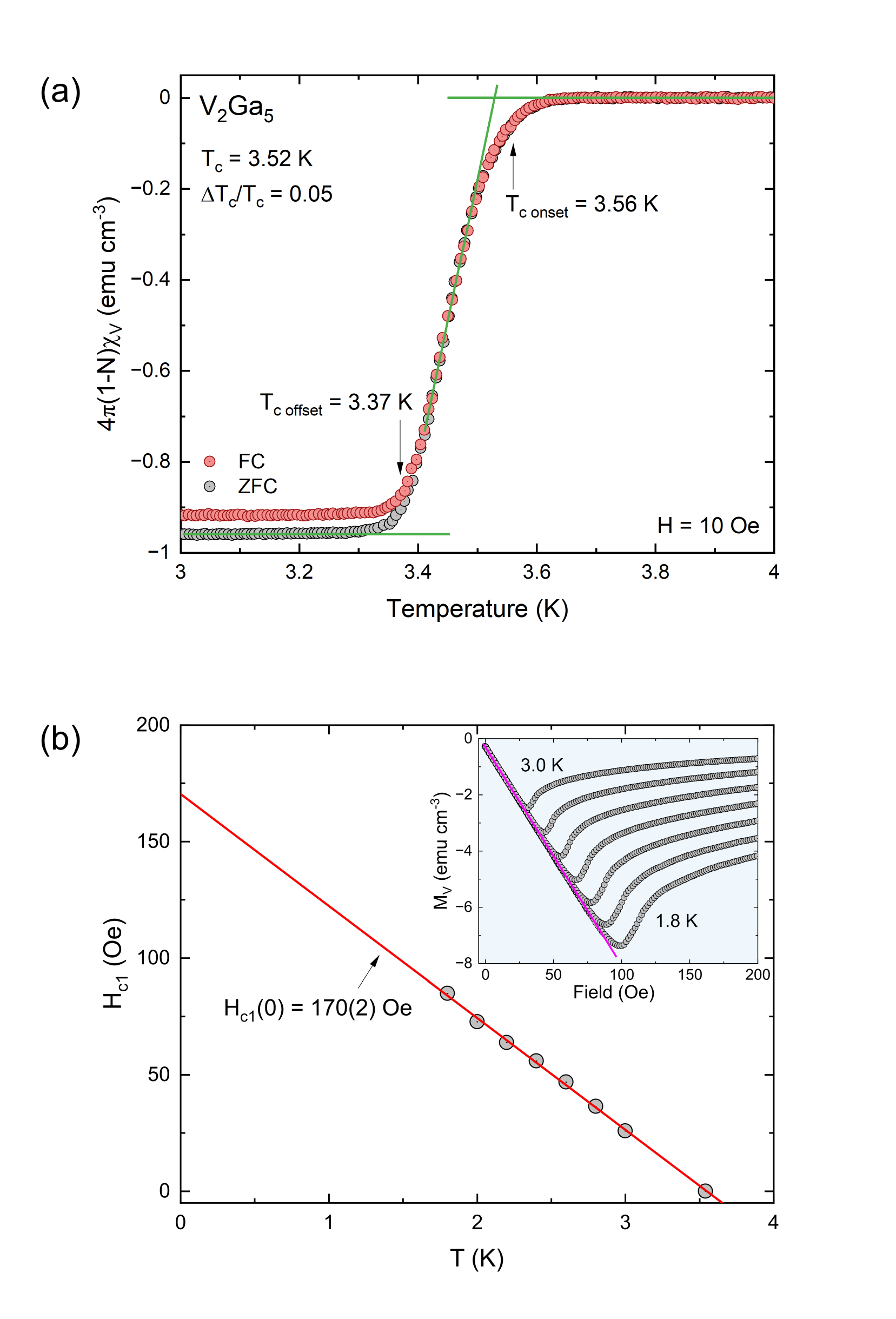}
\caption{(a) Zero-field-cooled (black) and field-cooled (red) magnetic susceptibility of \ce{V2Ga5} single crystal, corrected for the demagnetization factor, measured in field H = 10 Oe parallel to the c-axis. T$_c$ onset and T$_c$ offset were calculated using the 5\%/95\% criterion, and T$_c$ was determined from the maximal-slope criterion (green). (b) Temperature dependence of the lower critical field H$_{c1}$ determined for H $\parallel$ c, along a linear fit to the data (red). In the inset, M$_V$(H) curves for several temperatures are presented. 
}
\label{Fig:ZFC}
\end{figure}

With the sample purity confirmed, we proceeded with magnetic measurements. Throughout this work, the magnetic susceptibility is approximated as $\chi \approx M/H$. The low-field (H = 10 Oe) magnetic susceptibility, shown in \Cref{Fig:ZFC}a, reveals a clear superconducting transition at \( T_c = \SI{3.52}{K} \), consistent with previous reports \cite{PRB-2024, PRR-2024, Rogacki-V2Ga5, V2Ga5-Hc1, scirep-2025}. The transition is sharp, with a relative width \( \Delta T_c / T_c = 0.05 \), and the small difference between zero-field-cooled (ZFC) and field-cooled (FC) curves indicates a high-quality single crystal. Below \( T_c \), ZFC $\chi$(T), after correction for the demagnetization effect, reaches \(-0.96\), close to the ideal value of \(-1\) expected for a perfect diamagnet \cite{Meissner1933, Klimczuk2023}. To estimate the demagnetization factor \( N \), the needle-shaped crystal was approximated as a cylinder, with \( N \) calculated using the expression proposed by Prozorov and Kogan \cite{Prozorov2018}:
\[
N^{-1} = 1 + 1.6 \frac{c}{a},
\]
where \( 2c \) is the cylinder height and \( 2a \) its diameter. For the \ce{V2Ga5} crystal studied in this work, \( N \approx 0.08 \), close to the theoretical limit \( N = 0 \) expected for an infinitely long cylinder aligned with the magnetic field.

\Cref{Fig:ZFC}b shows the temperature dependence of the lower critical field \( H_{c1} \) (corrected for demagnetization), determined as the point at which \( M(H) \) data start to deviate from linearity (the so-called M–M\textsubscript{fit} method). Typically, \( H_{c1}(T) \) displays negative curvature and follows the Ginzburg–Landau expression:
\[
H_{c1}(T) = H_{c1}(0) \left[1 - \left(\frac{T}{T_c}\right)^2 \right],
\]
where \( H_{c1}(0) \) is the extrapolated zero-temperature value. Here, \( H_{c1}(0) = \SI{170(2)}{Oe} \), and the temperature dependence shows minimal curvature, well approximated by a linear function. This behavior may result from the limited temperature range of the measurements, as the temperature dependence of \(1/\lambda^2\) measured down to \( T < \SI{0.5}{K} \) \cite{scirep-2025} exhibits a similar linear trend between \SI{1.8}{K} and \SI{4}{K}, followed by saturation at lower temperatures, consistent with a fully gapped superconducting state.

\pagebreak

\begin{figure*}[t]
\centering
\includegraphics[width=0.95\textwidth]{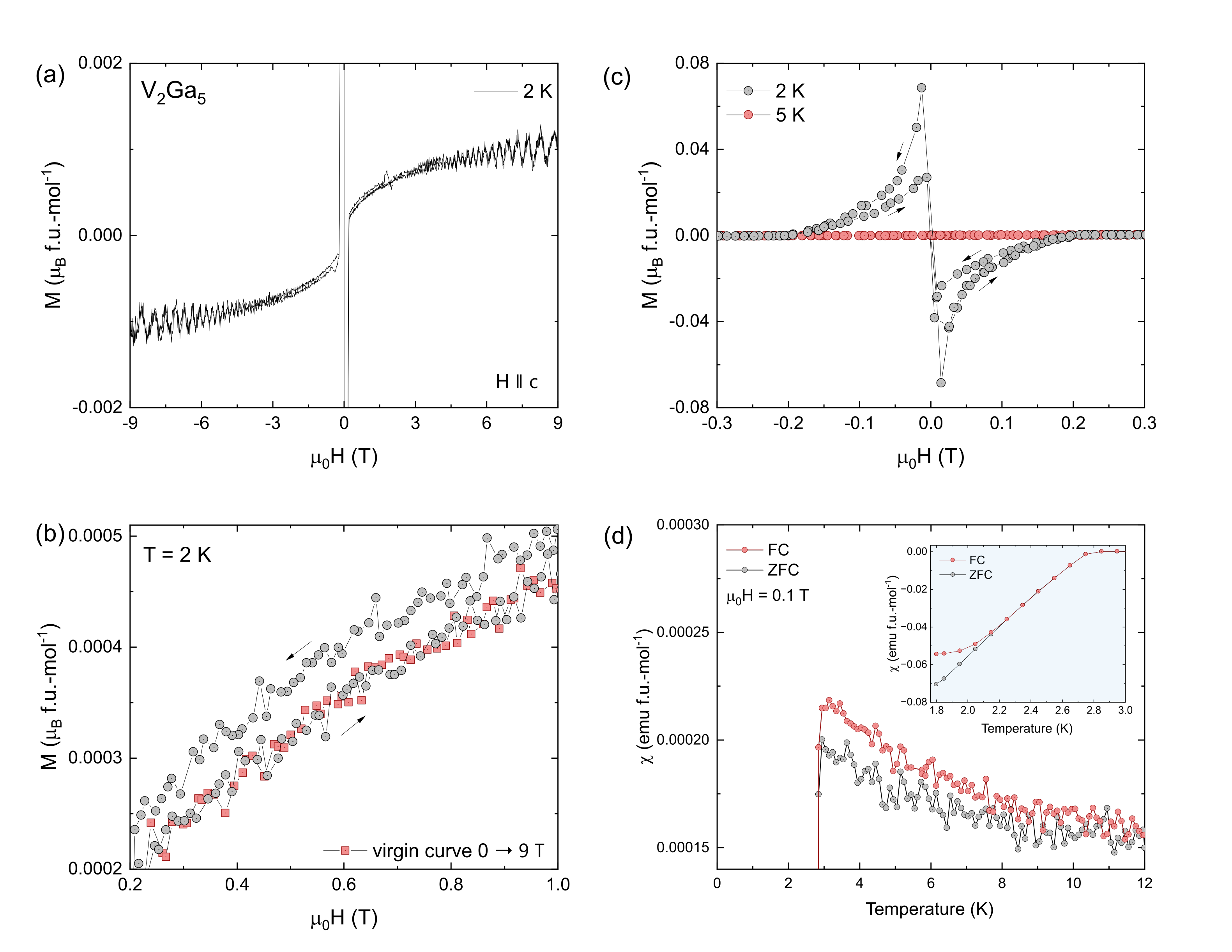}
\caption{ (a) $M(H)$ loop in the $\pm 9$ T field range measured at 2 K, showing clear high-field saturation and quantum oscillations. (b) Expanded view of the low-field $M(H)$ curve, revealing hysteresis that persists for $H>H_{c2}$. The bottom dataset is denser, as it is both the virgin curve (red squares) and the curve measured after the full $\pm 9$ T loop. (c) Low-field $M(H)$ highlighting the diamagnetism present at 2 K, which vanishes at 5 K, with hysteresis in the superconducting state. (d) Temperature dependence of magnetic susceptibility $\chi(T)$ for \ce{V2Ga5} single crystals, measured under field-cooled (FC, red) and zero-field-cooled (ZFC, black) conditions in the field $\mu_{0}H=\SI{0.1}{T}$. Clear ZFC/FC splitting is observed above $T_c$; inset shows the superconducting transition. All measurements were performed for $H \parallel c$.}
\label{Fig:MvsH}
\end{figure*}

We next performed detailed field-dependent magnetization measurements. The $M(H)$ loop measured over the $\pm \SI{9}{T}$ field range (\Cref{Fig:MvsH}a) exhibits both high-field saturation and pronounced de Haas–van Alphen (dHvA) oscillations, with the latter indicative of superior sample quality. Both features were previously reported \cite{PRB-2025}, where the dHvA oscillations were discussed in the context of topological properties. However, the strong orientation-dependent saturation, much more pronounced for $H \parallel c$, the direction of the vanadium chains, was not addressed. Our measurements therefore reproduce the main observations of Ref. \cite{PRB-2025}, extending them to both field sweep directions (marked with arrows in \Cref{Fig:MvsH}). An unexpected hysteresis is observed, most clearly visible in \Cref{Fig:MvsH}b, characteristic of ferromagnetic materials rather than conventional superconductors. The lower dataset in \Cref{Fig:MvsH}b corresponds to both the initial curve (virgin; red squares) and the measurement repeated after completing the full $\pm \SI{9}{T}$ loop, with the two overlapping perfectly, demonstrating that the hysteresis is not an experimental artifact.  The observed hysteresis occurs above the upper critical field \( \mu_0 H_{c2}(2\,\mathrm{K}) \approx \SI{0.3}{T} \), and differs from the hysteresis present in the superconducting state (\Cref{Fig:MvsH}c). Furthermore, as shown in \Cref{Fig:MvsH}d, the temperature-dependent magnetic susceptibility measured in the field of \( \mu_0 H = \SI{0.1}{T} \) shows an upturn below $T \approx \SI{10}{K}$, accompanied by a splitting between the ZFC and FC curves.

To confirm that the behavior is intrinsic to \ce{V2Ga5} and not sample-specific, we repeated the measurements on a polycrystalline specimen, whose single-phase nature was verified by powder X-ray diffraction (Figure S2). \noindent Compared with the thin rod-like single crystals, the larger volume of the polycrystalline sample provided a higher signal-to-noise ratio. Magnetization measurements on the polycrystal (Figure S3) closely matched those of the single crystal, suggesting the intrinsic origin of the observed behavior. The full saturation seen in \(M(H)\) for the \ce{V2Ga5} single crystal (\(H \parallel c\), \Cref{Fig:MvsH}a), together with the partial S-shaped saturation in the polycrystalline sample (Figure S3a) and in the single crystal with \(H \perp c\) \cite{PRB-2025}, indicate that saturation of \(M(H)\) is linked to the quasi-1D vanadium chains running along the c-axis. This anisotropy in the physical properties is also evident when comparing the superconducting \( M(H) \) curves (\Cref{Fig:MvsH}c and S3c). Although \ce{V2Ga5} is unambiguously a type-II superconductor, the single crystal displays an unusual re-entrance of diamagnetism, a “type-I-like” behavior, during field sweeps (see arrows in \Cref{Fig:MvsH}c), whereas the polycrystalline sample exhibits significant hysteresis, as expected for a type-II superconductor. This difference may arise from the near-zero pinning force, as reported for the \( H \parallel c \) configuration in single-crystalline \ce{V2Ga5} \cite{scirep-2025}, consistent with the geometry used in \Cref{Fig:MvsH}c. In contrast, the higher defect density in polycrystals provides effective pinning centers, producing a conventional type-II \( M(H) \) response (Figure S3c).

\pagebreak
The weak hysteresis and saturation in M(H), as well as the ZFC/FC splitting observed below $T \approx 10$ K in the absence of a clear transition in M(T), are typical of paramagnetic compounds with emerging ferromagnetic correlations, i.e. in the regime $T \gtrsim T_C$. The putative long range order does not develop, however, as it is suppressed by the onset of superconductivity at $\mathrm{T_c = 3.52}$ K, consistent with recent muon spin rotation (\(\mu_{SR}\)) and nuclear magnetic resonance (NMR) measurements, which detected no static internal magnetic fields \cite{PRR-2024, scirep-2025}, neither in the superconducting nor in the normal state. Indeed, if intrinsic, the emerging magnetism in \ce{V2Ga5} is extremely weak, as evidenced by the tiny saturation moment observed in this work (\Cref{Fig:MvsH}a), \( \mu_{\mathrm{sat}} \approx 0.001\,\mu_B/\text{f.u.} \) Such small moments are not unprecedented and have been reported in systems on the verge of magnetic order \cite{YbNi4P2, UBe13, YbRh2Si2}. For example, in \ce{YbNi4(P_{1-x}As_x)2} with \( x = 0.08 \), \(\mu_{SR}\) spectra also show absence of static magnetic fields, with the ordered moment limited to \( \mu_{\mathrm{ord}} < 0.005\,\mu_B \) \cite{YbNi4P2}, below the experimental resolution. Alternatively, the observed properties could be extrinsic, due to a small amount of a ferromagnetic impurity phase. Assuming a saturated moment of 1 $\mu_B$ per impurity atom, the observed saturation moment \( \mu_{\mathrm{sat}} \approx 0.001\,\mu_B/\text{f.u.} \) corresponds to impurity concentration $\mathrm{\approx}$ 0.1\%, well below the resolution of powder X-ray diffraction. Nonetheless, the observed behavior is independent of the synthesis method used (flux growth/arc-melting), and in the following we will present additional evidence supporting its intrinsic nature.

\subsection{Electrical resistivity}
\begin{figure}[htbp]
\centering \includegraphics[width=0.99\columnwidth]{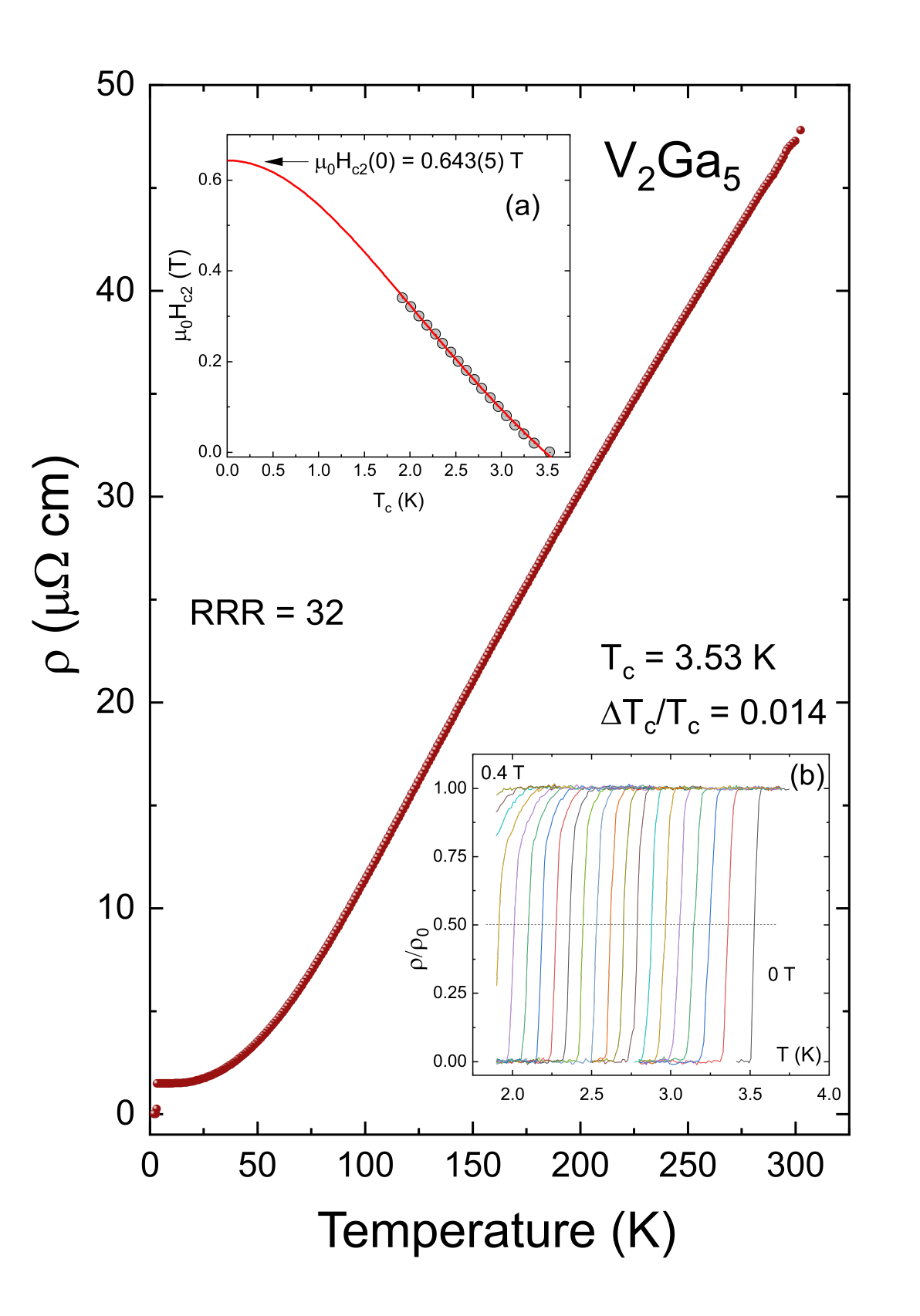}
\caption{Temperature dependence of the resistivity for a \ce{V2Ga5} single crystal in the 1.9 - 300 K temperature range, showing metallic behavior with a high residual resistivity ratio $RRR = 32$. Inset (a) presents the temperature dependence of the upper critical field H$_{c2}$, determined from the midpoint of the resistive transition under applied magnetic field. The extrapolated zero-temperature value is $\mu_{0}H_{c2}(0) = 0.643(5)$ T. Inset (b) shows normalized resistivity $\rho/\rho_{0}$ as a function of temperature under various magnetic fields, illustrating the systematic suppression of the superconducting transition with increasing field. The transition remains sharp, with minimal field-induced broadening up to $\mu_0$H = 0.4 T. 
}
\label{Fig:rho-all-temps}
\end{figure}

The electrical resistivity of \ce{V2Ga5} single crystal, measured in the 1.9 - 300 K temperature range is shown in \Cref{Fig:rho-all-temps}. The sample exhibits metallic behavior, with a residual resistivity ratio (RRR) of 32, similar to previous studies \cite{PRB-2024, PRB-2025, PRR-2024, V2Ga5-Fermi, Rogacki-V2Ga5}. The insets display 
the evolution of the upper critical field \( H_{c2} \), determined using the \( \rho(T_c) = 0.5\,\rho_0 \) criterion (a) and the superconducting transition under applied magnetic fields (b). The temperature dependence of \( H_{c2} \) was analyzed using the Ginzburg–Landau model:
\[
\mu_0 H_{c2}(T) = \mu_0 H_{c2}(0) \left[ \frac{1 - t^2}{1 + t^2} \right],
\] where \( t = T / T_c \) is the reduced temperature. The extrapolated zero-temperature value is \( \mu_0 H_{c2}(0) = \SI{0.643(5)}{T} \) and \( T_c = \SI{3.48(1)}{K} \), very similar to a recent report \cite{PRR-2024}.

\begin{figure}[htbp]
\centering
\includegraphics[width=0.99\columnwidth]{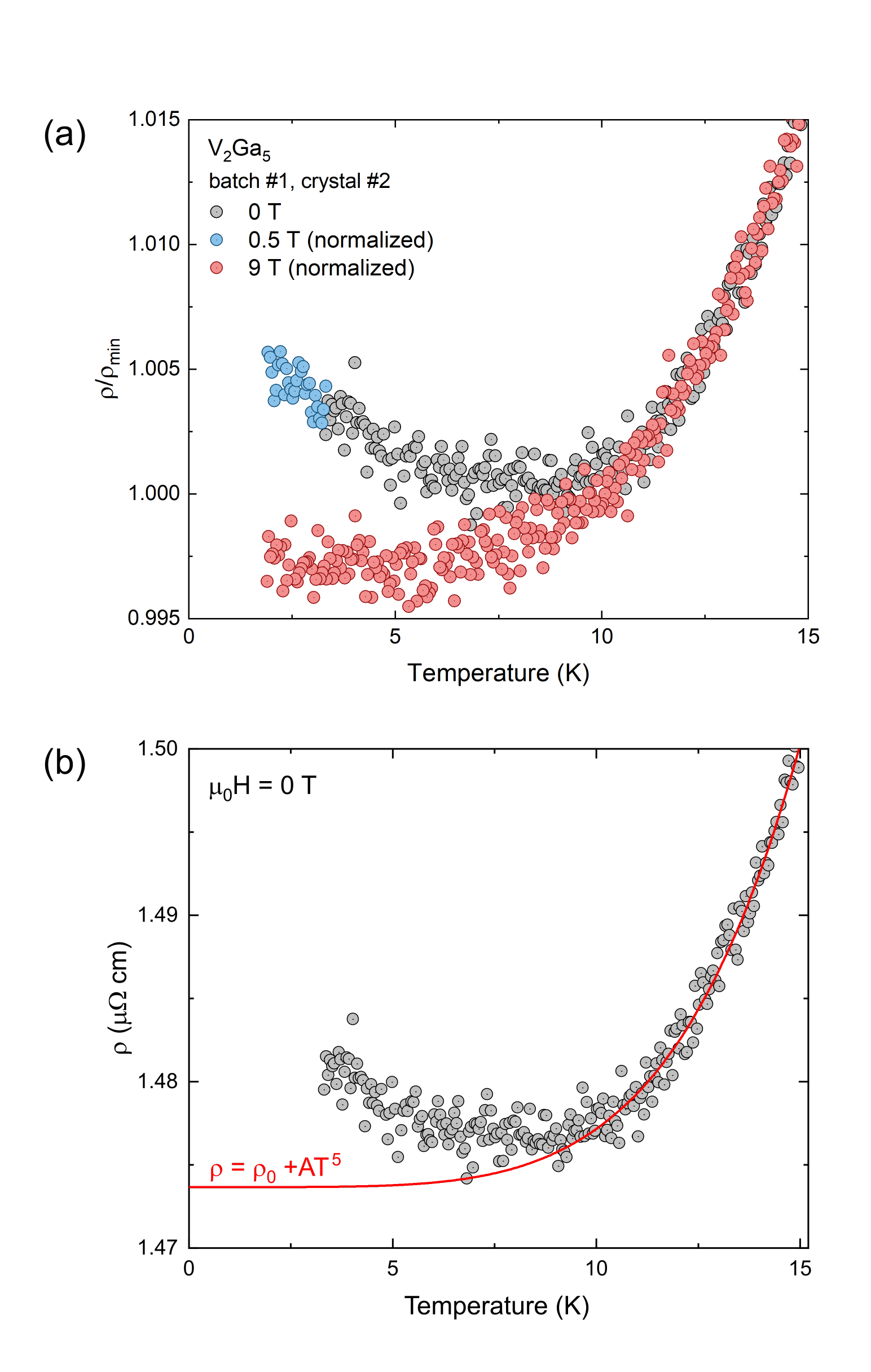}
\caption{(a) Resistivity of \ce{V2Ga5} batch \#1, crystal \#2 (see text for details) at 0 T, 0.5 T, and 9 T, showing an upturn below $T \approx 10$ K, enhanced by a small field of 0.5 T and suppressed by the 9 T field. The 0.5 T and 9 T datasets are shifted downwards to correct for the positive magnetoresistance. (b) Resistivity of the same crystal, with a $\mathrm{\rho(T) = \rho_0 + AT^5}$ fit performed in the 10 - 15 K temperature range presented in red.}  
\label{Fig:rho-upturn}
\end{figure}

To investigate whether the anomalous properties observed in the magnetic susceptibility measurements are reflected in electrical transport, we performed detailed low-temperature resistivity measurements. As shown in \Cref{Fig:rho-upturn}, an upturn in resistivity appears below $T \approx 10$ K, coinciding with the temperature at which the ZFC and FC magnetic susceptibility ($\chi$) curves begin to diverge. While all other measurements presented in this work were performed on crystals from batch \#1, here we also examined specimens from additional growth batches (Figures S4 and S5). These measurements confirm that the observed behavior is reproducible and not sample-specific; all samples exhibited the resistivity upturn, although the magnitude of the effect varied. We further examined whether this variation is related to impurity content, sample quality and/or dimensions by comparing the values of $T_c$, RRR, geometric factor S/l, and residual resistivity $\rho_0$ across all studied samples (Table S3). The magnitude of the upturn appears to be significantly correlated with the sample dimensions. To isolate this variable, we performed further measurements (Figure S6), with detailed discussion provided in the Supplementary Information. We found that the upturn is extremely fragile, amounting to only a fraction of a percent, and its reliable detection therefore requires both meticulous sample preparation (very thin and long crystals) and high-precision resistivity measurements, which likely explains why it was not identified in previous studies.

The above observations strongly suggest that the resistivity upturn is an intrinsic property of \ce{V2Ga5}. Electron-electron interactions \cite{Lee1985} cannot account for the upturn, which shows a pronounced field dependence. 
The Kondo effect \cite{Kondo1964, Krolak2025} would lead to a field-suppressed resistivity upturn, but it is incompatible with the diverging ZFC/FC magnetic susceptibility, the presence of hysteresis in M(H), and the magnetic-field dependence of the specific heat discussed in the next section. The upturn could also be caused by weak localization \cite{Lee1985} or structural inhomogeneity \cite{witek2024, Jaroszynski2008, vanDelft2020}. Both scenarios, however, appear irrelevant, as the single crystals are sufficiently clean to exhibit clear dHvA oscillations, with a relatively high $\mathrm{RRR} = 32$ and no sign of inhomogeneity was detected with EDX analysis. Moreover, preliminary measurements on Mn-doped \ce{V2Ga5} show that the upturn is the strongest in the undoped sample and is progressively suppressed, eventually disappearing with increasing Mn content, while the $\mathrm{RRR}$ decreases precipitously at the same time, suggesting an anticorrelation between the upturn and disorder. A more detailed investigation of the physical properties of \ce{V_{2-x}Mn_xGa5} is currently underway.

Given the hysteresis and saturation observed in \(M(H)\), the low-temperature upturn in \(\rho(T)\) might be attributed to emerging ferromagnetic spin fluctuations. According to Fisher–Langer theory \cite{Fisher-Langer}, for \(T \gtrsim T_C\) the magnetic contribution to the resistivity of a ferromagnetic metal, \(\rho_{\mathrm{m}}\), develops a maximum and then decreases rapidly for \(T < T_C\) as spin disorder scattering is suppressed. In most ferromagnets this contribution appears only as a kink in \(\rho(T)\) \cite{Krolak2022, NdIr3, Kaczorowski1990, Gofryk2006}, being masked by the other terms, i.e. the Fermi-liquid and lattice contributions. To quantify the nonmagnetic background in \ce{V2Ga5}, we fitted the resistivity in the 10–15 K range, shown in \Cref{Fig:rho-upturn}b. In this temperature window the data are well described by \(\rho(T) = \rho_0 + A T^5\), i.e., by the standard \(T^5\) dependence from electron–phonon scattering, with the \(T^2\) term being negligible. Because the phonon contribution falls off steeply below T \(\approx 10\) K, the temperature dependence of resistivity in this temperature range can be ascribed essentially entirely to the magnetic contribution, \(\rho_{\mathrm{m}}(T)\), attributed to spin fluctuations. Thus, the observed increase in resistivity is naturally explained by the Fisher–Langer theory \cite{Fisher-Langer}. This hypothesis is consistent with the magnetic–field-induced suppression of the upturn; while the total magnetoresistance (MR) is substantial and positive \cite{PRB-2025}, after normalizing the 0 T and 9 T curves at T = 15 K to account for the positive MR (\Cref{Fig:rho-upturn}a), the low-temperature upturn is suppressed by the magnetic field, as typically observed for scattering on spin fluctuations. Additional support for this hypothesis comes from the pressure dependence of the resistivity in \ce{V2Ga5}. As shown in Figure 4(a) of Ref. \cite{PRB-2025}, \(\rho(T)\) decreases with pressure for \(T \gtrsim \SI{20}{K}\) but increases below \(T \approx \SI{6}{K}\), implying that different scattering mechanisms dominate in these two temperature regimes. Interestingly, the low-temperature increase of \(\rho\) with pressure closely resembles the increase observed here under a small applied magnetic field (\(\mu_0 H = \SI{0.5}{T}\); see \Cref{Fig:rho-upturn}a, S4, and S5). Moreover, the positive magnetoresistance reported in Ref.~\cite{PRB-2025} is smaller at \(T = \SI{5}{K}\) than at 10 K and 20 K, in contrast to the usual increase in MR with decreasing temperature \cite{Ni3In7, Pavlosiuk2017, Pavlosiuk2018}. This trend is naturally explained if, below T \(\approx \SI{10}{K}\), there is an additional negative magnetoresistance arising from the suppression of spin fluctuation scattering. Thus, while the microscopic origin of the upturn requires further investigation, our measurements suggest that it is intrinsic in nature and potentially linked to the anomalies observed in magnetic susceptibility measurements.

\subsection{Specific heat}

\begin{figure}[htbp]
\centering
\includegraphics[width=0.99\columnwidth]{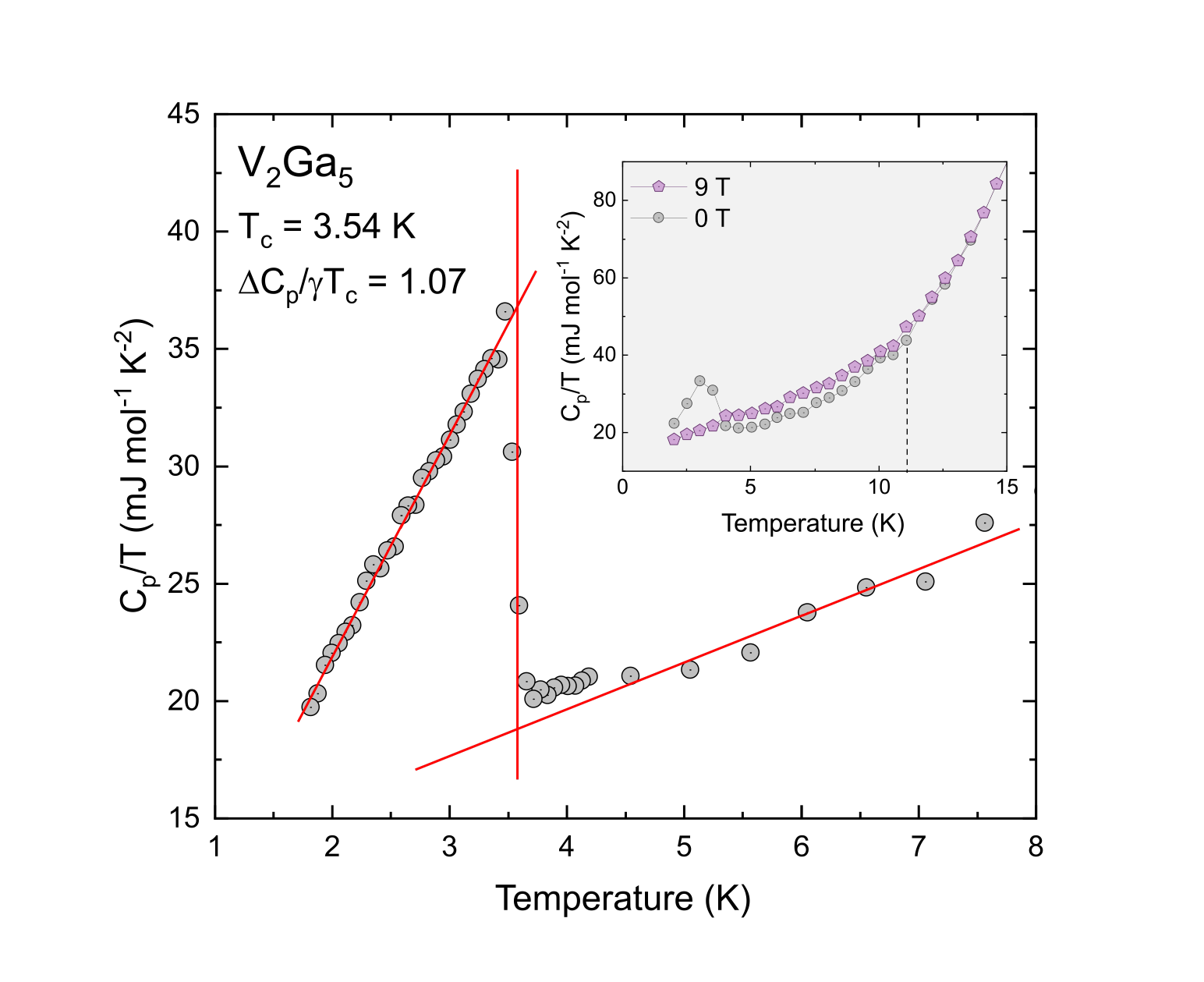}
\caption{Low-temperature $C_p/T$ vs $T$ showing a clear jump at $T_c = 3.54$ K. The estimated jump height yields $\Delta C_p/\gamma T_c = 1.07$. In the inset, $C_p/T$ vs $T$ measured in zero field (black) and in an applied field of 9 T (violet) are presented, with a field-induced enhancement of $C_p$ visible below $T \approx 11$ K.}
\label{fig:HC}
\end{figure}

\begin{figure}[htbp]
\centering
\includegraphics[width=0.80\columnwidth]{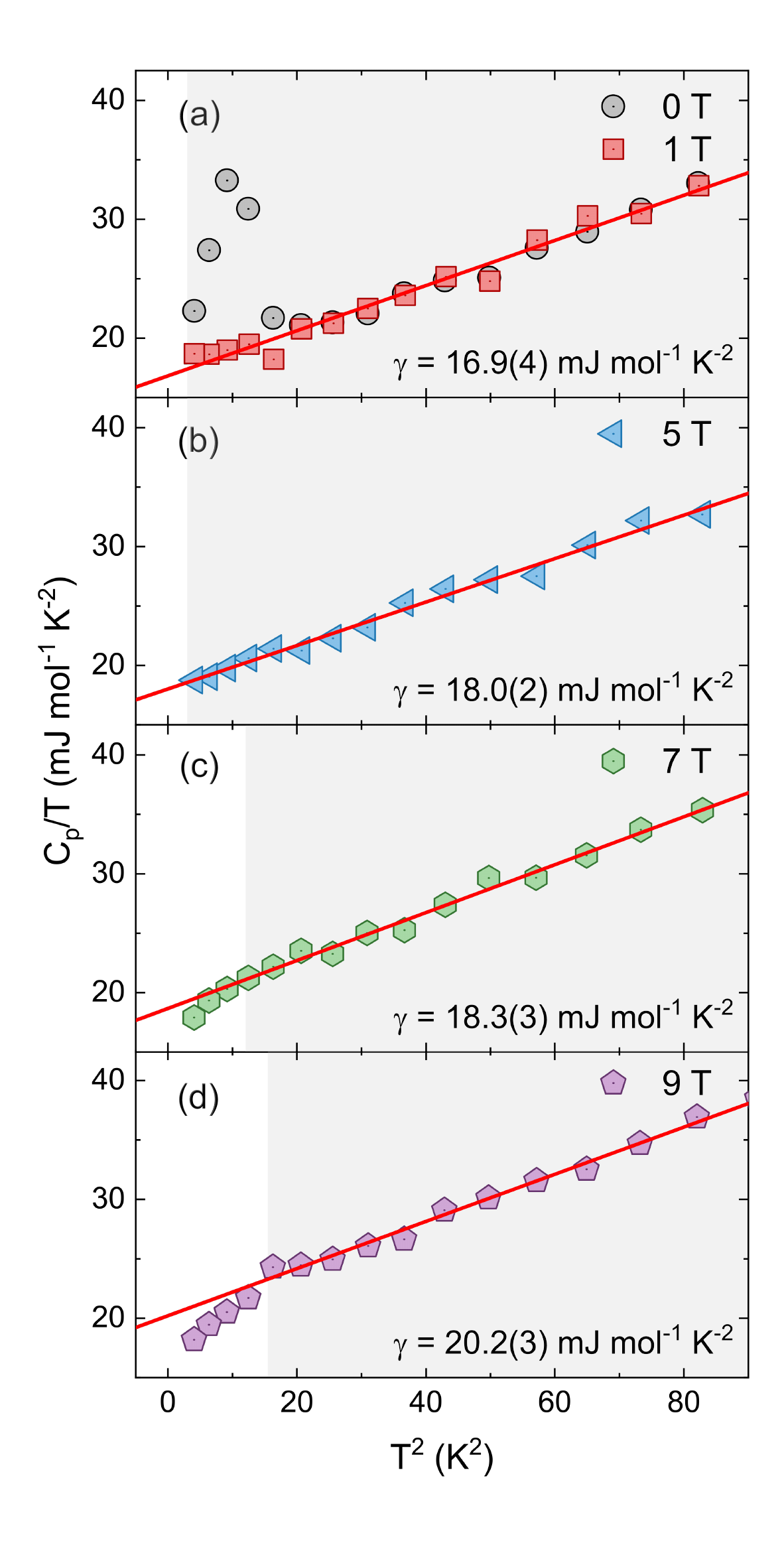}
\caption{Low-temperature specific heat of \ce{V2Ga5} plotted as $C_p/T$ versus $T^2$, measured under applied magnetic fields of (a) 0 and 1 T, (b) 5 T, (c) 7 T, and (d) 9 T. Solid red lines are linear fits to $C_p/T = \gamma + \beta T^2$ formula, with the fit range indicated by the shaded gray regions.}
\label{fig:HC-fields}
\end{figure}

To establish the bulk nature of superconductivity in \ce{V2Ga5}, heat-capacity measurements were performed. As shown in \Cref{fig:HC}, a sharp superconducting transition occurs at $T_c = \SI{3.54}{K}$, consistent with the magnetic susceptibility and electrical resistivity data. The Sommerfeld coefficient, $\gamma = \SI{16.9(4)}{mJ\,mol^{-1}\,K^{-2}}$, was obtained from a linear fit to the $C_p/T$ versus $T^2$ data measured under an applied magnetic field of $\mu_0 H = \SI{1}{T}$ (\Cref{fig:HC-fields}a). The resulting normalized heat capacity jump, $\Delta C_p / \gamma T_c = 1.07$, is smaller than the BCS prediction for a single-gap $s$-wave superconductor ($\Delta C_p / \gamma T_c = 1.43$), in agreement with previous reports \cite{PRB-2024, PRR-2024, V2Ga5-Fermi}. This behavior was previously attributed to the presence of two superconducting gaps, as inferred from low-temperature specific heat measurements \cite{PRB-2024}. While a reduced jump is often explained by multigap superconductivity, it may also result from gap anisotropy \cite{Clem1966}, or from a combination of the two. A detailed analysis of the \ce{V2Ga5} gap anisotropy is currently underway. 

The inset of \Cref{fig:HC} shows the specific heat of \ce{V2Ga5} measured in zero field and under an applied magnetic field of \( \mu_0 H = \SI{9}{T} \). To the best of our knowledge, the specific heat of this compound has not previously been reported under such a high magnetic field for \( T > T_c \). An unexpected magnetic-field dependence emerges below \( T \approx \SI{11}{K} \), coinciding with the anomalies observed in magnetic susceptibility and electrical resistivity measurements. To better visualize the field-induced enhancement, measurements at additional magnetic fields were performed, as shown in \Cref{fig:HC-fields}. For the magnetic field range 0-5 T, the field only enhances the Sommerfeld coefficient. For \( \mu_0 H > \SI{5}{T} \), the low-temperature curvature also changes, with the effect becoming more pronounced at higher magnetic fields.

A small amount of a ferromagnetic impurity phase on the order of 0.1\% could easily remain undetectable in \(\mathrm{C_{p}/T}\) measurements and therefore cannot account for the observed field-induced enhancement. Moreover, even if such an impurity were detectable, \(\mathrm{C_{p}/T}\) of a ferromagnet is expected to decrease with increasing magnetic field \cite{Pfleiderer2001}. Likewise, in the case of a dilute Kondo effect, an applied magnetic field is also expected to suppress the specific heat \cite{Satoh1989}. Thus, following the analysis presented in the preceding sections, we ascribe the observed field dependence to ferromagnetic spin fluctuations. Indeed, a similarly shaped low-temperature curvature in $\mathrm{C_p/T}$ vs $\mathrm{T^2}$ was observed in \ce{Sr(Co_{1-x}Ni_{x})2As2} for $x = 0.20$ \cite{Sangeetha2019}, and was found to grow for $x = 0.30$ upon approaching a quantum critical point. This interpretation would imply that \ce{V2Ga5} is driven closer to ferromagnetism by the application of a magnetic field. To confirm this scenario, tuning study of \ce{V2Ga5} (e.g. with Mn substitution) is required to determine whether a similar enhancement of the low-temperature upturn can be achieved. 

\newpage
The observed field-induced enhancement of specific heat is consistent with what is generally expected for ferromagnetic spin fluctuations. In the $T < T_C$ regime, an applied magnetic field is expected to suppress spin fluctuations, resulting in a decrease of the specific heat \cite{Pfleiderer2001}. In contrast, in the $T \gtrsim T_C$ regime, the specific heat is often enhanced by an applied magnetic field. A representative example is \ce{Sc_{3.1}In} \cite{Sc3In-Ames,Svanidze2015}, where the specific heat above the Curie temperature is enhanced by magnetic fields up to approximately \SI{5}{T}; only at higher fields are spin fluctuations effectively quenched. Similarly, in \ce{Sr(Co_{1-x}Ni_{x})2As2} \cite{Sangeetha2019} ($x = 0.25$ and $0.30$), as well as in \ce{Ca(Co_{1-x}Ni_{x})_{2-y}As2} \cite{Pakhira2021} ($x = 0.16$--$0.52$), the low-temperature upturn attributed to ferromagnetic spin fluctuations is strongly suppressed by magnetic field, while its high-temperature tail is actually enhanced.

\begin{figure*}[t]
\centering
\includegraphics[width=0.99\textwidth]{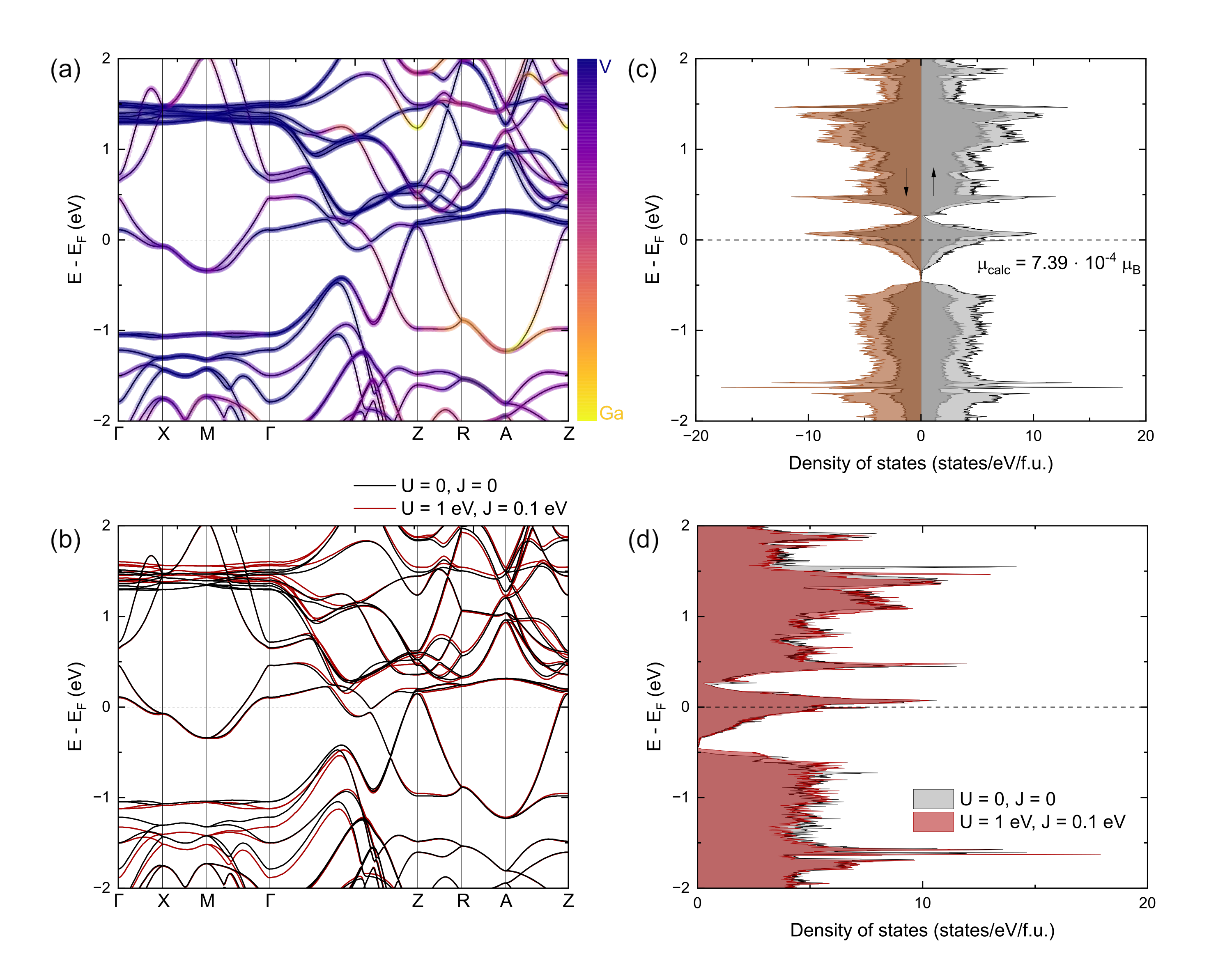}
\caption{(a) Atom-projected electronic band structure of \ce{V2Ga5}, with the color scale indicating the relative contributions from V and Ga states. (b) Comparison of electronic structures calculated without (black) and with on-site Hubbard repulsion (U) and intra-atomic exchange interaction (J) corrections (red). (c) Spin-resolved density of states (DOS) for the U = 0, J = 0 calculation, yielding a  magnetic moment $\mu_{\mathrm{calc}}=7.39\cdot10^{-4} \ \mu_{\mathrm{B}}$ per unit cell. The dark shaded regions represents the contribution from V-\emph{d} states. (d) Majority-spin DOS for $U=0$, $J=0$ (gray) and $U=\SI{1}{eV}$, $J=\SI{0.1}{eV}$ (red).}
\label{Fig:DFT}
\end{figure*}

\pagebreak Based on the measurements presented in this paper, we argue that \ce{V2Ga5} resides in the $T \gtrsim T_C$ regime, where ferromagnetic correlations are only beginning to develop. This is consistent with the small ZFC/FC splitting, the tiny saturation moment, and the weak hysteresis observed in $M(H)$, and possibly also with the low-temperature upturn in the electrical resistivity. Then, as explained above, the enhancement of the specific heat in an applied magnetic field is an expected feature, in agreement with the results presented here. In the following section, we present arguments for a theoretical driving force behind proximity to ferromagnetic order in \ce{V2Ga5}.

\subsection{Electronic structure calculations}

To understand why ferromagnetism may be a favorable ground state for \ce{V2Ga5}, spin-polarized electronic structure calculations were performed, presented in \Cref{Fig:DFT}. The calculated band structure is consistent with previous reports \cite{PRR-2024, PRB-2025, PRB-2024}, showing V-dominated quasi-one-dimensional bands in the \emph{ab} plane along $\Gamma$–X–M, which show appreciable dispersion and stronger V-Ga mixing along the $\Gamma$–Z direction. The spin-polarized density of states (DOS), shown in \Cref{Fig:DFT}c, exhibits only a very small difference between the majority (gray) and minority (brown) channels, corresponding to a calculated magnetic moment of $\mu_{\mathrm{calc}} = 7.39 \cdot 10^{-4}\,\mu_B$ per unit cell. Although small, this value is close to the experimentally observed saturation moment, $\mu_{\mathrm{sat}} \approx 10^{-3}\,\mu_B/f.u.$

\begin{table}[H]
\caption{Value of the magnetic moment per unit cell for different calculation parameters.}
\label{Tab:DFT}
\begin{ruledtabular}
\begin{tabular}{cccc}
$U$ (eV) & $J$ (eV)  & Magnetic moment per unit cell ($\mathrm{\mu_B}$)\\
\hline
0   & 0     & $7.39\times10^{-4}$ \\
0.5 & 0     & $7.55\times10^{-4}$ \\
1.0 & 0     & $5.47\times10^{-4}$ \\
1.0 & 0.1   & $1.25\times10^{-3}$ \\
1.5 & 0.1   & $4.50\times10^{-3}$ \\
\end{tabular}
\end{ruledtabular}
\end{table}

Such a small magnetic moment could in principle arise from a fortuitous placement of the Fermi energy level at a local DOS maximum. Thus, we repeated the calculations within the DFT+$U$ framework to account for electronic correlations among the V-\emph{d} electrons, which could shift the relative position of the Fermi level and modify the DOS, (de)stabilizing the magnetic order. A comparison of the electronic structure and DOS obtained without and with correlation corrections is shown in \Cref{Fig:DFT}b and \Cref{Fig:DFT}d, respectively. The inclusion of correlation corrections primarily affects the quasi-one-dimensional V-\emph{d} states, leaving the states in the vicinity of the Fermi energy level almost unchanged. A summary of calculations performed for different values of $U$ and $J$ is given in \Cref{Tab:DFT}. In all cases, the calculated magnetic moments remain close to the experimentally observed value.

Based on these results, we conclude that, without including the possibility of superconducting pairing, ferromagnetism is the preferred ground state of \ce{V2Ga5}. This finding supports the interpretation of the experimental results presented in this manuscript. The origin of this tendency toward ferromagnetism can be understood within the Stoner framework \cite{Stoner}. As shown in the present calculations, the Fermi energy of \ce{V2Ga5} lies on a pronounced peak in the density of states. One requirement for itinerant ferromagnetism, i.e. a relatively high $\mathrm{DOS}(E_F)$ is therefore satisfied. The second requirement is a sufficiently large Stoner parameter $I_F$. Generally, $I_F$ is enhanced for more localized electronic states, increasing across the 3\emph{d} series with the increasing nuclear charge, and similarly decreasing in a column, from 3\emph{d} to 5\emph{d} elements as orbital delocalization increases \cite{Sigalas1994}. As discussed in the Introduction, the V-derived states close to the Fermi energy level have predominantly V–V antibonding character \cite{Rogacki-V2Ga5}. Dronskowski and Landrum \cite{Landrum2000} previously identified antibonding states as a chemical driving force for ferromagnetism; this relationship was later linked to positive effective exchange constants calculated using multiple-scattering theory \cite{Samolyuk2007}. In fact, antibonding states are inherently more localized than bonding states due to their increased nodal structure \cite{Sterling2024}. This enhanced localization provides a natural mechanism for an increased $I_F$, thereby favoring ferromagnetism.

\subsection{Summary} 
\pagebreak
Ferromagnetism and superconductivity, while both mechanisms for removing electronic instabilities from the Fermi surface, are typically antagonistic, with one suppressing the other. In this work, we presented a careful investigation of single- and polycrystalline \ce{V2Ga5}, establishing its phase and chemical purity and demonstrating that the observed phenomena are intrinsic and reproducible across multiple batches and samples. Based on the following phenomena observed below T $\approx$ 10 K: (1) ZFC/FC splitting in $\chi(T)$, accompanied by saturation and hysteresis in $M(H)$, (2) a magnetic-field-dependent upturn in resistivity, (3) an enhancement of the low-temperature specific heat in applied magnetic field, and (4) electronic structure calculations indicating a ferromagnetic ground state, we conclude that ferromagnetic correlations likely develop in \ce{V2Ga5} around 10 K. No single impurity- or defect-based mechanism can consistently account for the correlated anomalies observed in magnetization, electrical transport, and heat capacity measurements. The absence of long-range magnetic order suggests that ferromagnetic correlations are suppressed by the onset of superconductivity. This behavior, rare among intermetallic compounds, together with the high sample purity and the absence of intrinsic magnetic elements, identifies \ce{V2Ga5} as a promising platform for studying the interplay between superconductivity and weak itinerant ferromagnetism in a \emph{d}-electron system. 
Further studies, including electron doping, are essential to determine whether quantum critical behavior can be observed and to directly probe the symmetry of the superconducting
order parameter (e.g., via STM or point-contact spectroscopy).

\begin{acknowledgments}
S.K. is grateful to Peter Samuely for stimulating discussion. Work done on Gdańsk Tech was supported by the Platinum Joining Gdańsk Tech Research Community project (2/2/2023/IDUB/I.1B/Pt). X.H. and T.T.T. acknowledge the Arnold and Mabel Beckman Foundation, the NSF CAREER award NSF-DMR-2338014, and the Camille Henry Dreyfus Foundation. S.K. used ChatGPT-5 for grammar correction, and the AI-generated text was carefully reviewed afterward.
\end{acknowledgments}

\section*{Data availability}
The datasets used and/or analyzed during the current study are available from the corresponding authors upon reasonable request.

\nocite{*}

\bibliography{reference}

\end{document}